\documentclass[aps,preprint,epsfig]{revtex4-1}
\usepackage{amsmath,amssymb,graphicx}
\usepackage{latexsym}
\usepackage{esint}
\usepackage{color}
\usepackage{float}
\usepackage{multirow}
\usepackage{hyperref}

\def\ba{\begin{eqnarray}}
\def\ea{\end{eqnarray}}
\def\be{\begin{equation}}
\def\ee{\end{equation}}

\def\bm{\begin{math}}
\def\me{\end{math}}

\newcommand{\red}{\textcolor{red}}

\begin{document}

\title{Enhancement in Breaking of Time-reversal Invariance in the Quantum Kicked Rotor}
\thispagestyle{empty}

\author{Ramgopal Agrawal}
\email{ramgopal.sps@gmail.com}
\author{Akhilesh Pandey}
\email{apandey2006@gmail.com}
\author{Sanjay Puri}
\email{purijnu@gmail.com}
\affiliation{School of Physical Sciences, Jawaharlal Nehru University, New Delhi -- 110067, India.}

\begin{abstract}
We study the breaking of \textit{time-reversal invariance} (TRI) by the application of a magnetic field in the \textit{quantum kicked rotor} (QKR), using Izrailev's finite-dimensional model. There is a continuous crossover from TRI to time-reversal non-invariance (TRNI) in the spectral and eigenvector fluctuations of the QKR. We show that the properties of this TRI $\rightarrow$ TRNI transition depend on $\alpha^2/N$, where $\alpha$ is the chaos parameter of the QKR and $N$ is the dimensionality of the evolution operator matrix. For $\alpha^2/N \gtrsim N$, the transition coincides with that in random matrix theory. For $\alpha^2/N < N$, the transition shows a marked deviation from random matrix theory. Further, the speed of this transition as a function of the magnetic field is significantly enhanced as $\alpha^2/N$ decreases.\\
\red{(To appear in Physical Review E.)}
\end{abstract}

\maketitle

\section{Introduction}

The \textit{quantum kicked rotor} (QKR) has been a prototype model for quantum chaos. It was first introduced by Casati et al.~\cite{Casati1979} as a quantum analog of the \textit{classical kicked rotor} (CKR). An important paradigm of quantum chaos states that the spectral properties of quantum chaotic systems should follow the predictions of \textit{random matrix theory} (RMT)~\cite{PhysRevLett.52.1,RevModPhys.53.385,MEHTA2004,Haake2010,RevModPhys.69.731,GUHR1998189}. For time-independent systems, Gaussian ensembles of random matrices are appropriate. On the other hand, for time-periodic systems, circular random matrix ensembles are more useful. Izrailev~\cite{PhysRevLett.56.541,IZRAILEV1990299} introduced a model for the evolution operator of the QKR in $N$-dimensional Hilbert space, and found that the quasi-energy fluctuations agree with RMT. The discrete symmetries present in the QKR classify the spectral statistics according to the universality classes of RMT. In case the QKR maintains \textit{time-reversal invariance} (TRI), the statistics is described by \textit{circular orthogonal ensembles} (COE). The TRI is broken by applying a strong magnetic field, and the spectral statistics then becomes consistent with that of \textit{circular unitary ensembles} (CUE). However, at moderate perturbation strengths, it is found that the spectral fluctuations exhibit intermediate statistics, and the TRI is said to be partly broken~\cite{Pandey1993,PhysRevLett.67.1}.

Over the decades, breaking of TRI in quantum chaotic systems has received special attention in the literature~\cite{RevModPhys.69.731,PhysRevB.43.14390,Bohigas_1995,PLUHAR19951}. According to the \textit{von Neumann-Wigner theorem}, TRI-breaking is accompanied by increased level repulsion in the spectrum~\cite{MEHTA2004,Haake2010}. Berry and Robnik~\cite{Berry_1986} studied chaotic billiards with a single line of magnetic flux as a perturbation, and confirmed this level repulsion. The eigenvector properties also change during the breaking of TRI~\cite{ROSENZWEIG1968437,PhysRevLett.54.2313,PhysRevB.50.11267,PhysRevLett.85.2482}. When TRI is preserved, the absolute-squared eigenvector component has a $\chi^2_1$-distribution. This becomes a $\chi^2_2$-distribution when TRI is completely broken~\cite{RevModPhys.53.385,ROSENZWEIG1968437}. (We will provide the functional forms of the $\chi^2_1$- and $\chi^2_2$-distributions later.) In the crossover regime, the distribution is intermediate to $\chi^2_1$ and $\chi^2_2$. This TRI to \textit{time-reversal non-invariance} (TRNI) transition is driven by a control parameter (say $\lambda$), which measures the strength of the applied field. The influence of $\lambda$ depends on the dimensionality $N$. In the semi-classical limit ($N \rightarrow \infty$), the transition is discontinuous in $\lambda$. One identifies a {\it local transition parameter}, which smoothens the transition for systems with large $N$. This parameter is further useful in comparing the transition in different systems. It depends on the near-diagonal matrix elements of the TRI-breaking perturbation and the mean level spacing of the unperturbed spectrum~\cite{PhysRevLett.54.2313,FRENCH1988198}. In the context of classical RMT, Pandey and Mehta~\cite{Mehta_1983,Pandey1983} have investigated the spectral transition from \textit{Gaussian orthogonal ensembles} (GOE) to \textit{Gaussian unitary ensembles} (GUE). Using Dyson's Brownian motion model~\cite{MEHTA2004,Dyson1962}, they studied the continuous crossover from GOE $\rightarrow$ GUE. Later, Pandey and Shukla~\cite{Pandey_1991,Pandey2004} extended the idea of Brownian motion to unitary ensembles and studied the COE $\rightarrow$ CUE transition.

In the present paper, we investigate the effect of the TRI $\rightarrow$ TRNI transition on the spectral and eigenvector fluctuations of the QKR. We report results from a detailed numerical study of this transition. The motivation for this work comes from our recent proposal of \textit{finite-range Coulomb gas (FRCG) models} for random matrix ensembles, and their applicability to a finite-dimensional matrix model of the QKR~\cite{PhysRevE.96.052211,PhysRevE.101.022217,PhysRevE.101.022218}. The CKR is characterized by a parameter $\alpha$, with the system being highly chaotic for $\alpha \gg 1$. The relevant parameter in the QKR is $\alpha^2/N$, where $N$ is the size of the evolution operator matrix~\cite{PhysRevLett.64.5,PhysRevE.96.052211}. In our recent work \cite{PhysRevE.96.052211,PhysRevE.101.022217,PhysRevE.101.022218}, we have proposed that the spectral statistics of the QKR with chaos parameter $\alpha$ is described by an FRCG model of range $d = \alpha^2/N$. These FRCG models are generalizations of the Dyson's Coulomb gas model to the case where eigenvalues have a restricted interaction range.

Let us next consider the TRI $\rightarrow$ TRNI transition in the QKR. For $\alpha^2/N \gtrsim N$, this is described well by the COE $\rightarrow$ CUE transition of RMT~\cite{Pandey1993,Shukla_1997,PhysRevE.53.1362}. This is because the TRI-breaking perturbation, when represented in the diagonal basis of the TRI-preserving part, belongs to the Gaussian RMT ensembles. However, when $\alpha^2/N < N$, the perturbation is no longer completely random. There has been almost no investigation of this crossover when $\alpha^2/N < N$. This is the open problem which we address in this paper. As we will show later, the perturbation matrix takes the form of a \textit{power-law random band matrix} (PRBM)~\cite{PhysRevE.54.3221,MIRLIN2000259}, whose bandwidth depends on the parameter $\alpha^2/N$. The significant perturbation matrix elements lie inside the band, and determine the nature of the transition.

The main observations of our study are as follows: (a) For small values of $\alpha^2/N$, the TRI $\rightarrow$ TRNI transition in the spectral fluctuations of QKR does not obey the usual COE $\rightarrow$ CUE transition. Further, the speed of the transition is enhanced with the decrease in $\alpha^2/N$. (b) The transition in one-point eigenvector statistics also does not obey the COE $\rightarrow$ CUE transition for any fixed value of $\alpha^2/N < N$. As in the case of spectral fluctuations, the transition is enhanced with decrease in $\alpha^2/N$.

This paper is organized as follows. In Sec.~\ref{s2}, we discuss TRI-breaking transitions in RMT using a Brownian motion model and introduce the local transition parameter. In Sec.~\ref{s3}, we briefly review the QKR and discuss its spectral properties. In Sec.~\ref{s4}, we study the matrix elements of the TRI-breaking perturbation in the QKR. The transition in spectral fluctuations of the QKR is studied in Sec.~\ref{s5}. In Sec.~\ref{s6}, the one-point eigenvector fluctuations are analyzed. Finally, in Sec.~\ref{s7}, we summarize this paper and discuss our results.

\section{\label{s2} Breaking of time-reversal invariance: Local Transition Parameter}

The TRI-breaking transition in time-periodic quantum systems is driven by an external magnetic field of strength $\lambda$. The time-evolution operator of the system can be expressed as
\be
\label{eq1}
\mathcal{U}(\lambda) = \mathcal{U}_0~{\rm e}^{i \lambda V},
\ee
where $\mathcal{U}_0$ is the unperturbed part and $V$ is the TRI-breaking perturbation.

It is known that quantum systems possessing TRI and/or exact symmetries are often modeled by the stationary ensembles of classical RMT. In such systems, the partial violation of TRI and/or an exact symmetry results in a crossover between local fluctuation properties. The system, during this crossover, follows intermediate Brownian motion ensembles (BME)~\cite{Dyson1962,Pandey_1991}. In the context of quantum maps, which we study in this paper, the relevant ensembles are the unitary BME. These are parameter-dependent ensembles of $N$-dimensional unitary matrices $U(\tau)$, where $\tau$ plays the role of fictitious time. The evolution of $U(\tau)$ under Brownian motion is described as follows:
\be
\label{eq2}
U(\tau+\delta\tau) = U(\tau)~{\rm e}^{i \sqrt{\delta \tau}~V(\tau)}.
\ee
Here, $\delta\tau$ is the infinitesimal time increment. At each $\tau$, a random perturbation is introduced via an $N$-dimensional Hermitian random matrix $V(\tau)$. In Dyson's model for circular ensembles, the perturbation $V(\tau)$, when represented in $U(\tau)$-diagonal basis, belongs to Gaussian ensembles of random matrices. On the other hand, in FRCG models, $V(\tau)$ takes the form of band matrices~\cite{PhysRevE.96.052211,PhysRevE.101.022217,PhysRevE.101.022218}. The Brownian motion of matrix elements in Eq.~\eqref{eq2} implies that the quasi-energies of the parameter-dependent ensembles execute Brownian motion under two-body Coulomb gas repulsion.

Under a nonlinear rescaling between the fictitious time $\tau$ and the TRI-breaking parameter $\lambda$ with $\tau = \lambda^2$ and $\delta \tau = (\delta \lambda)^2$, Eq.~\eqref{eq2} is appropriate to model TRI-breaking in the quantum system. As stated earlier, the transition in terms of $\lambda$ is dependent on the dimension $N$ and is discontinuous for $N \rightarrow \infty$~\cite{Berry_1986,Mehta_1983}. To obtain a smooth transition, one has to parameterize the transition in terms of a local transition parameter. This parameter is obtained from perturbation theory~\cite{PhysRevLett.54.2313,FRENCH1988198} as follows:
\be
\label{eq3}
\Lambda = \frac{\lambda^2~v^2}{D^2}.
\ee
Here, $D$ is the mean level spacing of the quasi-energy spectrum in the unperturbed case. Further, $v^2$ denotes the variance of near-diagonal matrix elements of the TRI-breaking perturbation $V(\lambda)$ in the diagonal basis of the TRI-part. The transition in nearest-neighbor fluctuations is complete when $\lambda v$ is nearly equal to the mean level spacing $D$, i.e., $\Lambda \sim \mathcal{O}(1)$. Eq.~\eqref{eq3} demonstrates the strong dependence of the transition on the perturbation matrix elements.

\section{\label{s3} Quantum Kicked Rotor}

In the presence of a magnetic field of strength $\lambda$, the Hamiltonian for the CKR is given by
\be
\label{eq4}
H = \frac{(p-\lambda)^2}{2} + \alpha \cos(\theta + \theta_0) \sum_{n=-\infty}^{\infty} \delta(t-nT),
\ee
where $\theta$ and $p$ are position and momentum variables. The parameter $\alpha$ is \textit{the kicking strength}, which introduces non-integrability in the system. Therefore, it is known as the \textit{chaos parameter}. The period between two consecutive kicks is denoted as $T$. The time-reversal ($p \rightarrow -p, t \rightarrow -t$) and parity ($\theta \rightarrow -\theta, p \rightarrow -p$) invariance are broken by parameters $\lambda$ and $\theta_0$, respectively.

The corresponding quantum dynamics can be understood in terms of a quantum map, known as the QKR-map, which describes the evolution of a state $|\psi_t \rangle$ as follows:
\be
\label{eq5}
\vert \psi_{t+T} \rangle = \mathcal{U} \vert \psi_t \rangle.
\ee
Here, $\mathcal{U}$ is the evolution operator defined over one period of time. It can be written as $\mathcal{U} = BG$, where
\be
\label{eqB}
B = \exp \left[- \frac{i\alpha \cos(\theta+\theta_0)}{\hbar} \right] 
\ee
describes instantaneous kicking. Further,
\be
\label{eqG}
G = \exp\left[-\frac{i T (p-\lambda)^2}{2\hbar} \right]
\ee
corresponds to the free motion of the rotor between two consecutive kicks. The position and momentum variables $\theta$ and $p$ are now operators.

When dealing with TRI-breaking in quantum maps, the symmetric form of $\mathcal{U}$ is more convenient to work with. This is related to other representations by a unitary transformation, i.e.,
\be
\label{eq6}
\mathcal{U} = B^{1/2}GB^{1/2}.
\ee
The chosen form of $\mathcal{U}$ is manifestly TRI at $\lambda=0$. For any non-zero $\lambda$, $\mathcal{U}$ is TRNI. Thus, the momentum operator $p$ is the TRI-breaking perturbation in the system. If we write $\mathcal{U}$ in a basis in which $B$ is diagonal, i.e., position basis (see Appendix A for details), we obtain the matrix elements:
\be
\label{eqInf}
\begin{split}
\mathcal{U}_{qq^{\prime}} = &\exp\left[-i\frac{\alpha}{2 \hbar} \left\{\cos\left(\theta_q+\theta_0\right) + \cos\left(\theta_q^{\prime}+\theta_0\right)\right\} \right] \\
&\times\sum_{l=-\infty}^{\infty} \exp\left[-i \frac{T}{2 \hbar}(l \hbar - \lambda)^2\right] \exp\Big[i~l(\theta_q-\theta_q^{\prime})\Big].
\end{split}
\ee

Since the momentum basis for the QKR is unbounded, the eigenvalues of the infinite matrix $\mathcal{U}$ depend on the continuous parameters $\theta_q$ and $\theta_q^{\prime}$. For a discrete spectrum \cite{PhysRevE.87.062905,IZRAILEV1990299}, one is often interested in the level fluctuations. However, there are infinite quasi-energies in the interval $(0,2\pi)$. Therefore, the mean level spacing is zero, and the mean level density is infinite, i.e., the fluctuations are Poisson. To observe level repulsion analogous to that in bounded time-independent quantum chaotic systems \cite{PhysRevLett.52.1,Haake2010}, we need a finite-dimensional model of the QKR. Such a model was first formulated by Izrailev~\cite{PhysRevLett.56.541}. (We will discuss later that, even in a finite-dimensional model, the spectral fluctuations can be Poisson.) In the Izrailev model, an additional periodic boundary condition is imposed on the momentum basis, i.e., the corresponding classical phase space becomes bounded in both position and momentum \cite{*[{Regardless of whether the phase space of the CKR is cylindrical (bounded in one variable) or toroidal (bounded in both variables), its statistical properties at large kicking strengths are unchanged. See~}][{}] ott1993chaos}. (There also exist other possible ways to approximate Eq.~\eqref{eqInf}, e.g., truncation of the infinite matrix $\mathcal{U}$~\cite{IZRAILEV1990299}.) In Izrailev's model, the representation of $\mathcal{U}$ is chosen in a generic basis with a finite number of levels (say, $N$). In the limit $N \rightarrow \infty$, this model approaches the original infinite-dimensional QKR in Eq.~\eqref{eqInf}. (See also the recent work of Manos-Robnik \cite{PhysRevE.87.062905,PhysRevE.91.042904}.)

To employ periodic boundary conditions in the momentum basis, we consider the momentum eigenstates to obey the following condition: $\vert l \rangle = \vert l + N \rangle$, where the label $l$ is an integer. This gives discrete values for the eigenvalues of $\theta$:
\be
\theta_m = \frac{2\pi m}{N}, ~~~~~ m = 0, \pm 1, \pm 2, \ldots.
\ee
Note that $\theta_m$ can take $N$ discrete values in the period $\theta \rightarrow \theta + 2\pi$. Now the position basis, as well as the momentum basis, are periodic and both have a finite number of basis vectors $N$. In this case, the evolution operator $\mathcal{U}$ in position basis is the $N$-dimensional matrix:
\be
\label{eq7}
\begin{split}
\mathcal{U}_{mn} = &\frac{1}{N} \exp\left[-i\frac{\alpha}{2} \left\{\cos\left(\frac{2\pi m}{N}+\theta_0\right) + \cos\left(\frac{2\pi n}{N}+\theta_0\right)\right\} \right] \\
&\times\sum_{l=-N_1}^{N_1} \exp\left[-i\left(\frac{l^2}{2}-\lambda l-2\pi l~\frac{(m-n)}{N}\right)\right].
\end{split}
\ee
Here, $N_1 = (N-1)/2$, and the indices $m,n = -N_1, -N_1+1,\ldots, N_1-1,N_1$. Without loss of generality, we have set $T = \hbar = 1$. For $N \rightarrow \infty$, the model in Eq.~\eqref{eq7} converges to the infinite-dimensional model in Eq.~\eqref{eqInf} (see Appendix A).

For the CKR, we know that the dynamics depends entirely on $\alpha$. If $\alpha\gg1$, the stroboscopic map of the CKR is fully chaotic. The spectral properties of the corresponding quantum map are expected to follow RMT predictions~\cite{PhysRevLett.52.1}, but this statement is only partly true. As discussed above, the original QKR (which has an infinite-dimensional matrix for the time-evolution operator $\mathcal{U}$) has Poisson quasi-energy statistics for all $\alpha$. This occurs due to the localization of eigenstates of the evolution operator in the momentum basis. This phenomenon is known as \textit{dynamical localization}, and is analogous to \textit{Anderson localization} in the 1-$d$ infinite lattice with disorder \cite{PhysRevLett.49.509,PhysRev.109.1492}. The traces of dynamical localization also appear in the $N$-dimensional model of the QKR with finite but large $N$. This is analogous to Anderson localization in the 1-$d$ finite lattice~\cite{PhysRevLett.64.5}. Therefore, the quantum dynamics actually depends on two parameters: $\alpha$ and $N$. For applicability of RMT, apart from the requirement of chaotic classical dynamics, the eigenstates of the evolution operator should also be fully extended in the momentum basis. This second requirement depends on the parameter $\alpha^2/N$~\cite{PhysRevE.96.052211,PhysRevLett.64.5}. When $\alpha^2/N \gtrsim N$, all the eigenstates are extended. However, when $\alpha^2/N$ decreases below $N$, the eigenstates start to become localized. This gives rise to level crossing in the spectrum, and the spectral statistics is then intermediate to Poisson and RMT.

We emphasize that the dynamics of the CKR is invariant under change in parameters $\lambda$ and $\theta_0$~\cite{Pandey1993,Shukla_1997}. However, the corresponding quantum dynamics is not invariant. The quantum dynamics has TRI for $\lambda = 0$, but this is destroyed for $\lambda \ne 0$. In the latter case, if $\theta_0 = 0$, TRI remains broken but an additional TP-invariance holds (TP is the product of time-reversal and parity operators). This TP-invariance of the QKR is broken by setting $\theta_0 \ne 0$. Further, when both $\lambda = 0$ and $\theta_0 = 0$, the parity operator commutes with the Hamiltonian and both can be diagonalized in a common basis. On the other hand, for $\theta_0 \ne 0$, the parity operator no longer commutes with the Hamiltonian.

The spectral properties of the evolution operator can be analyzed in terms of the {\it number variance statistics} $\Sigma^2(r)$, which is defined as the variance of the number of quasi-energies $n$ in an interval of size $rD$, where $D$ is the mean level spacing ($=2 \pi /N$):
\be
\label{eq8}
\Sigma^2(r) = \overline{n^2} - \overline{n}^2.
\ee
Here, the bar denotes the spectral or ensemble average. To numerically calculate $\Sigma^2(r)$ from Eq.~\eqref{eq8}, we first find the quasi-energy spectrum by diagonalizing $\mathcal{U}$ in Eq.~\eqref{eq7}. The quasi-energies are then unfolded using the mean level density $N/2 \pi$. We consider a range of values for $\alpha^2/N$, with $\alpha \gg 1$ and a fixed $N=2001$. Statistical averages are performed on an ensemble of $50$ independent spectra, obtained by locally varying $\alpha$ in the range $[\alpha-5,\alpha+5]$. This slight variation of $\alpha$ is adequate to ensure the statistical independence of the spectra. The central value of $\alpha$ will be used to label the corresponding spectra.

\begin{figure}[p]
\centering
\includegraphics[width=0.8\textwidth]{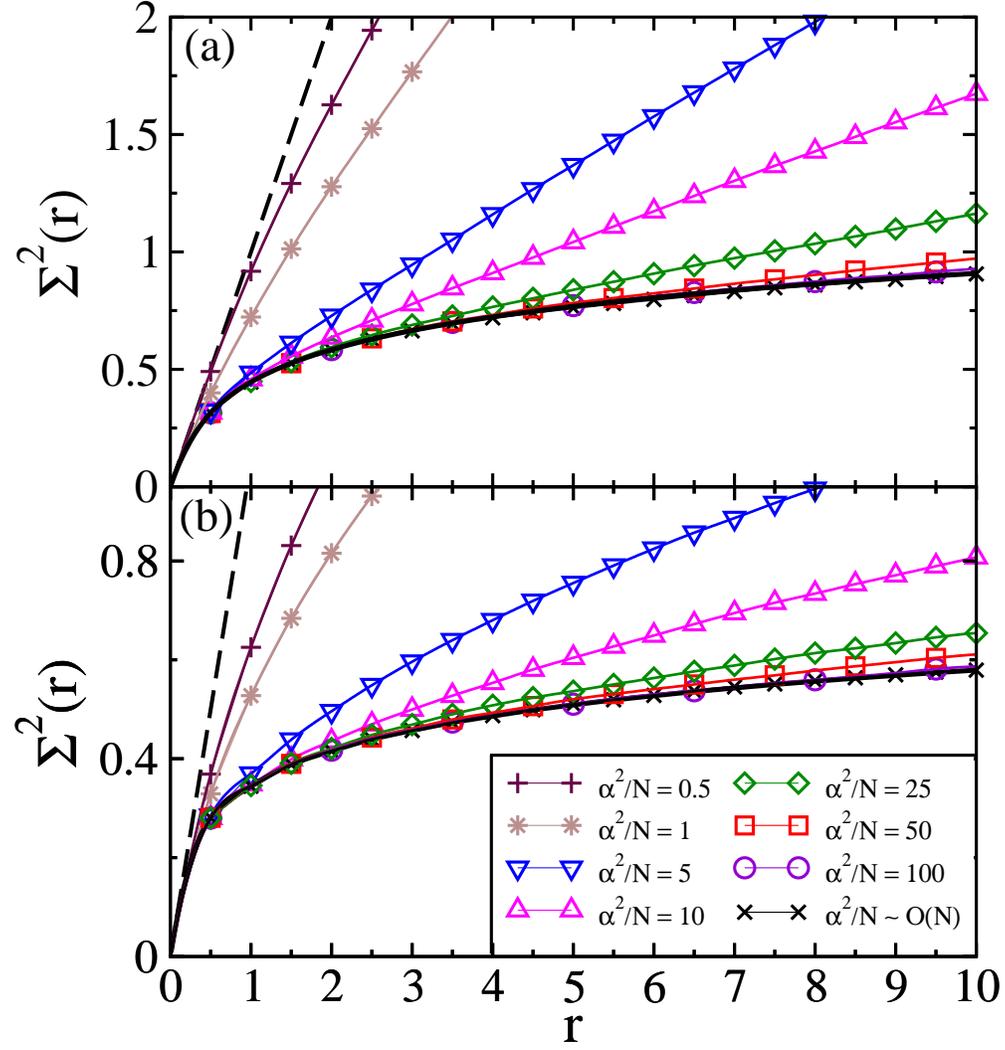}
\caption{Number variance $\Sigma^2(r)$ vs. $r$ for various values of $\alpha^2/N$ (with $N = 2001$) for (a) $\lambda = 0$, and (b) $\lambda = 0.9$. The dashed and solid curves denote the Poisson and RMT results, respectively.}
\label{fig1}
\end{figure}

We first discuss the spectral properties of QKR when TRI is preserved ($\lambda = 0$), and TRI is fully broken ($\lambda \simeq 1$). In both cases, $\theta_0$ is fixed to a nonzero value: $\theta_0 = \pi/(2N)$. In Fig.~\ref{fig1}, we plot $\Sigma^2(r)$ vs. $r$ for various values of $\alpha^2/N$. In Fig.~\ref{fig1}(a), we show the case with $\lambda=0$. Since TRI is maintained, one expects the number variance to follow COE predictions for $\alpha^2/N \sim \mathcal{O}(N)$~\cite{PhysRevLett.56.541}, which is seen in the figure. For small values of $\alpha^2/N < N$, one sees a clear deviation from COE in the direction of the Poisson result. As $\alpha^2/N$ is reduced, the deviation from COE is stronger at larger values of $r$. Note that, for any value of $\alpha^2/N < N$, $\Sigma^2(r)$ attains the RMT-limit (COE, in this case) in the interval size $r \ll \alpha^2/N$ and becomes Poisson-like ($\Sigma^2 \sim r$) for $r \gg \alpha^2/N$~\cite{PhysRevE.96.052211,PhysRevE.101.022218}.

In Fig.~\ref{fig1}(b), we show analogous results for $\lambda=0.9$. In this case, TRI is completely broken due to a strong magnetic field, and $\Sigma^2(r)$ vs. $r$ for $\alpha^2/N \sim \mathcal{O}(N)$ coincides with the CUE result. For $\alpha^2/N <N$, $\Sigma^2(r)$ is higher than the CUE value and approaches the Poisson result, similar to Fig.~\ref{fig1}(a). Comparing the data sets at a fixed value of $\alpha^2/N$ in Figs.~\ref{fig1}(a)-(b), we conclude that the spectrum is more rigid in the TRNI case.

\section{\label{s4} Perturbation matrix elements}

In the previous section, we discussed the spectral properties of the QKR for zero and strong field strengths. However, when TRI is broken by the application of moderate field strengths ($\lambda \ll 1$), the local fluctuations exhibit intermediate statistics. We need to calculate the local transition parameter $\Lambda$ (introduced in Sec.~\ref{s2}) to characterize the crossover obtained by varying field strength. It depends on the near-diagonal matrix elements of the TRI-breaking perturbation, which is the momentum operator $p$. For $\alpha^2/N \gtrsim N$, the off-diagonal elements of $p$ in $\mathcal{U}(\lambda=0)$-diagonal basis are Gaussian random variables with zero mean and uniform variance, i.e., $ip$ is a typical member of GOE~\cite{Pandey1993,Shukla_1997}. The variance of these matrix elements depends linearly on $N$, $v^2 \simeq \hbar^2 N/12$. Then, the transition parameter $\Lambda$ using Eq.~\eqref{eq3} is given by
\be
\label{eq9}
\Lambda = \frac{\lambda^2 N^3}{48 \pi^2},
\ee
where the mean level spacing of the quasi-energies of $\mathcal{U} (\lambda=0)$ is $D = 2\pi/N$, and $\hbar$ is set to unity.

The above form of $\Lambda$ is valid for $\alpha^2/N \gtrsim N$. When $\alpha^2/N$ decreases below $N$, the operator $p$ in the diagonal basis of $\mathcal{U}(\lambda=0)$ is not fully random. In fact, it has a banded structure with significant matrix elements lying inside a band around the diagonal. The matrix elements are Gaussian distributed with mean zero, but the variance depends on the distance from the diagonal.

\begin{figure}[p]
\centering
\includegraphics[width=0.4\textwidth]{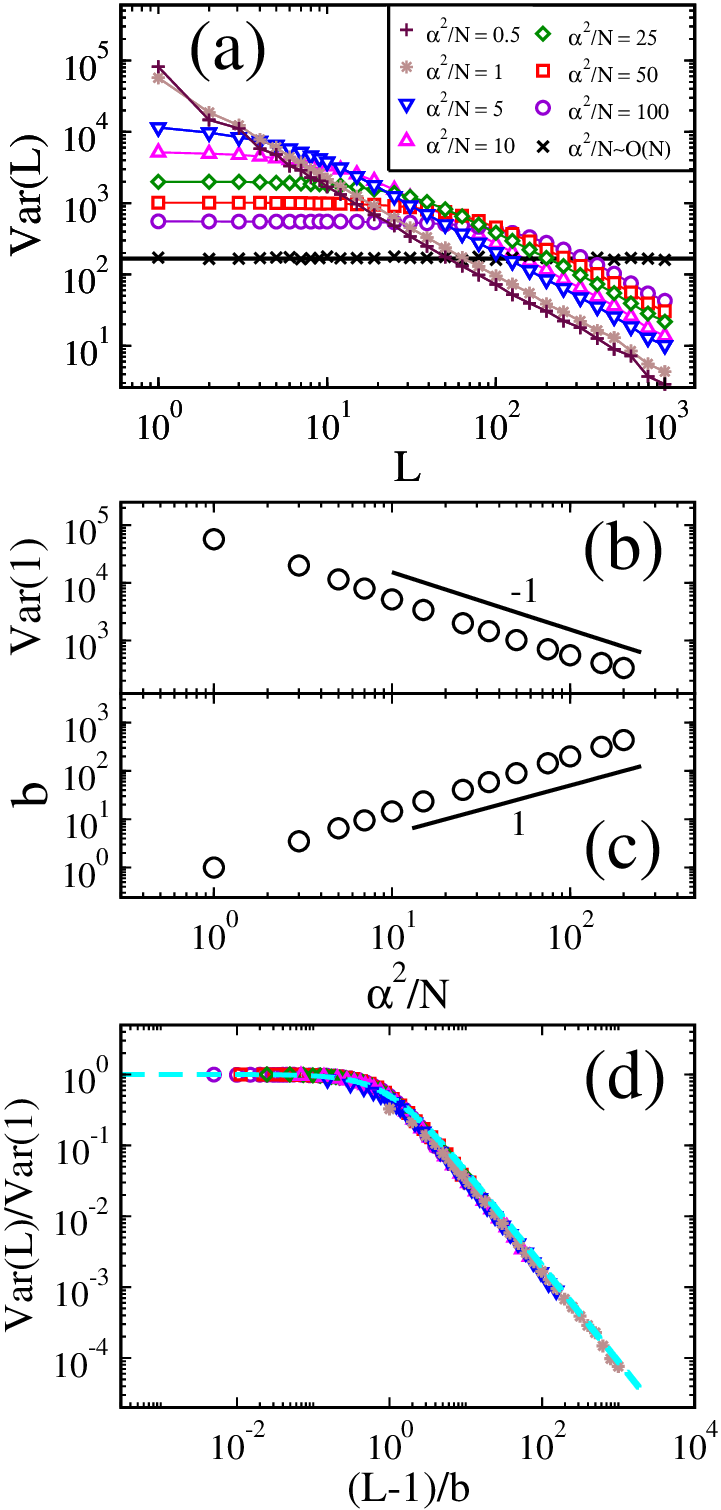}
\caption{(a) Plot of $Var(L)$ vs. $L$ on a log-log scale for different values of $\alpha^2/N$. The solid line denotes the semi-classical result: $Var(L) \simeq N/12$ with $N=2001$. (b) Plot of $Var(1)$ vs. $\alpha^2/N$ on a log-log scale. The solid line denotes the slope $-1$. (c) Plot of $b$ vs. $\alpha^2/N$ on a log-log scale. The solid line denotes the slope $+1$. (d) Plot of $Var(L)/Var(1)$ vs. $(L-1)/b$ on a log-log scale for the data sets in (a). The dashed curve is the best-fit to Eq.~(\ref{eq10}) with $m \simeq 1.35$.}
\label{fig2}
\end{figure}

In this section, we numerically investigate the perturbation matrix elements. For a value of $\alpha^2/N$ (with fixed $N$ = 2001), we first write the operator $p$ in $\mathcal{U}(\lambda=0)$-diagonal basis. The obtained matrix of $p$ is imaginary and anti-symmetric, as the diagonal basis of $\mathcal{U}$ in Eq.~\eqref{eq6} is time-reversal invariant at $\lambda=0$. To obtain good accuracy in the statistics, we construct an ensemble of 50 independent matrices of $p$ as described in Sec.~\ref{s3}. The matrix elements are denoted as $p_{ij}$. We have $\overline{\overline{p_{ij}}} = 0$, where the double bar denotes an ensemble average and an average over elements at the same distance $L=|i-j|$ from the diagonal. In Fig.~\ref{fig2}(a), we plot the variance of off-diagonal matrix elements $Var(L)$ vs. $L$ for different values of $\alpha^2/N$. For $\alpha^2/N \sim 1$, $Var(L)$ begins to fall as $L$ increases from $1$. For higher values of $\alpha^2/N$, $Var(L)$ is nearly constant inside a band around the diagonal with $|i-j| < b$, and decays for $|i-j| > b$. We define $b$ as the value of $L$ where $Var(L)$ decays to half of $Var(1)$. The bandwidth $b$ increases linearly with $\alpha^2/N$, and $Var(L=1) \sim N/\alpha^2$ [see Figs.~\ref{fig2}(b) and~\ref{fig2}(c)]. The fall of $Var(L)$ outside the band for a fixed value of $\alpha^2/N < N$ is linear on the log-log scale of the figure, showing that it has a power-law decay with $L$. (This power-law behavior will be further analyzed shortly.) Now, looking at data for $\alpha^2/N \sim \mathcal{O}(N)$ in the same figure, we see that the band has covered the entire matrix, $b \simeq N/2$. Therefore, the matrix $p$ becomes fully random with $Var(L)$ fluctuating around the theoretical value ($\simeq N/12$) for all $L$.

In Fig.~\ref{fig2}(d), we plot the scaled variables $y=Var(L)/Var(1)$ against $x=(L-1)/b$ on a log-log scale. This yields an excellent data collapse. The solid line denotes the best-fit to the collapsed data, and has the form
\be
\label{eq10}
y = \frac{1}{1+x^m},
\ee
with exponent $m \simeq 1.35$. At large $x$, Eq.~\eqref{eq10} has a power-law decay: $y\sim x^{-m}$. This peculiar behavior of $Var(L)$ shows that the TRI-breaking perturbation, when represented in $\mathcal{U}(\lambda=0)$-diagonal basis, belongs to a more general class of random matrices, known as \textit{power-law random band matrices} (PRBM)~\cite{PhysRevE.54.3221,MIRLIN2000259}. In PRBM, the variance of off-diagonal elements decays as a power-law with distance from the diagonal.

\begin{table}[t]
\caption{The variance $v^2$ of significant matrix elements around the diagonal for different values of $\alpha^2/N$ with $N$ = 2001.}
\label{tab1}
\begin{center}
		\begin{tabular}{| c | r |}
			\hline
			\hspace{3mm}$\alpha^2/N$\hspace{3mm} & \hspace{4mm} $v^2$ \hspace{4mm} \\
			\hline
			5 & 8227.92 \\
			\hline
			10 & 3822.59 \\
			\hline
			25 & 1515.06 \\
			\hline
			50 & 778.40 \\
			\hline
			100 & 428.63 \\ [0.5ex]
			\hline
		\end{tabular}
	\end{center}
\end{table}

We next discuss the main theme of this paper. Fig.~\ref{fig2}(b) shows that $Var(L)$ near the diagonal (at $L=1$) is higher for smaller values of $\alpha^2/N$. The reason behind this increase is the enhancement of overlaps between neighboring eigenstates of $\mathcal{U}(\lambda=0)$ with a decrease in $\alpha^2/N$. The matrix elements $p_{ij}$ fluctuate strongly when $p$ is represented in such highly correlated neighboring eigenstates. Therefore, the variance of $p_{ij}$ has higher values near the diagonal. One can see from Eq.~\eqref{eq3} that, even for a small perturbation strength $\lambda$, the transition parameter $\Lambda$ becomes large due to an increase in the variance. Therefore, the TRI $\rightarrow$ TRNI transition in terms of $\Lambda$ is enhanced for small values of $\alpha^2/N$. (We remind the reader that the TRI $\rightarrow$ TRNI transition only coincides with the COE $\rightarrow$ CUE transition in the limit $\alpha^2/N \sim N$.)

In Table~\ref{tab1}, we present values of variance $v^2$ for different $\alpha^2/N$. We compute $v^2$ as the variance of all elements of matrix $p$ up to the bandwidth $b$. These values will be used to obtain the transition parameter $\Lambda$ in the following discussion.

\section{\label{s5} Transition in Spectral Fluctuations}

When TRI-breaking takes place in the QKR, the resistance of levels to crossings in the spectrum becomes greater, i.e., the rigidity of the spectrum is increased. In the previous section, we have predicted an enhancement in the TRI $\rightarrow$ TRNI transition for small $\alpha^2/N$, which can be observed via a rapid increase in rigidity of the underlying spectrum. For this, we investigate the transition curves of the number variance $\Sigma^2(r)$, defined in Eq.~\eqref{eq8}, as a function of the transition parameter $\Lambda$ for different values of $\alpha^2/N$.

The TRI $\rightarrow$ TRNI transition curves for two-point fluctuations in Gaussian ensembles were obtained in~\cite{Mehta_1983,Pandey1983}. These are also valid for the COE $\rightarrow$ CUE transition in the large-$N$ limit~\cite{Pandey_1991}:
\be
\label{eq11}
\Sigma^2(r,\Lambda) = \Sigma^2(r,\infty)-2\int_{0}^{r}\mathrm{d}s~(r-s) \left[~Y_2(s,\Lambda)-Y_2(s,\infty)~\right].
\ee
Here, $Y_2(s,\Lambda)$ is the {\it two-level cluster function} for the transition ensemble:
\be
\label{eq12}
Y_2(s,\Lambda) = Y_2(s,\infty)-\left[\int_{0}^{1}\mathrm{d}x~\mathrm{e}^{2\pi^2\Lambda x^2} x~\sin(\pi x s) \int_{1}^{\infty}\mathrm{d}y~\mathrm{e}^{-2\pi^2\Lambda y^2}~\frac{\sin(\pi y s)}{y}\right].
\ee

\begin{figure}[p]
\centering
\includegraphics[width=0.8\textwidth]{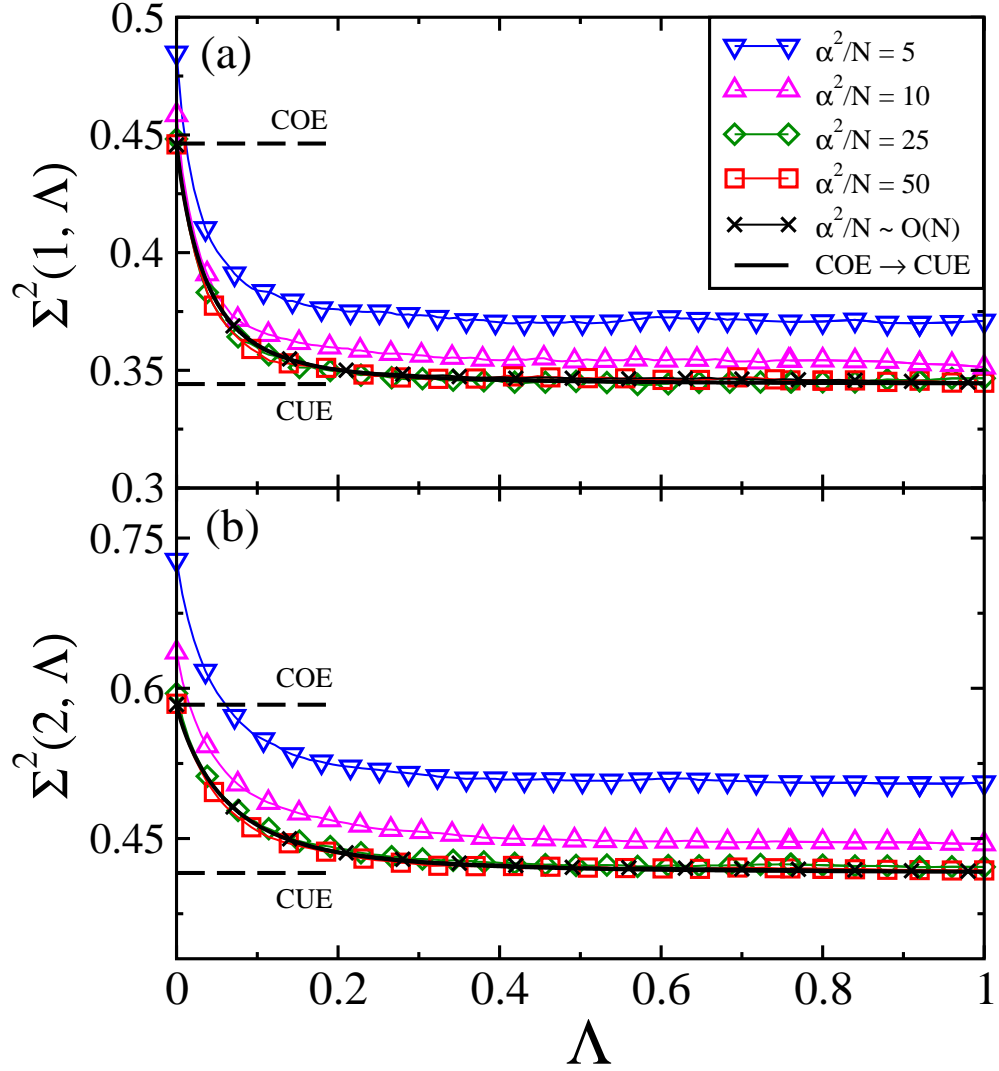}
\caption{Transition curves of number variance, $\Sigma^2(r,\Lambda)$ vs. $\Lambda$, for different values of $\alpha^2/N$ when (a) $r = 1$, and (b) $r = 2$. The parameter $\Lambda$ is estimated using Eq.~\eqref{eq3}. The solid curves are theoretical results for the COE $\rightarrow$ CUE transition. The dashed lines indicate the COE and CUE limits.}
\label{fig3}
\end{figure}

In Fig.~\ref{fig3}, we present numerical results for the transition curves of $\Sigma^2(r,\Lambda)$ vs. $\Lambda$ for various values of $\alpha^2/N$ in interval sizes $r=1$ and $2$. The transition parameter $\Lambda$ for different values of $\lambda$ is calculated from Eq.~\eqref{eq3}. The analytical result for the COE $\rightarrow$ CUE transition is obtained by numerically integrating Eqs.~\eqref{eq11}-\eqref{eq12}.

Let us first consider Fig.~\ref{fig3}(a) for $r=1$. When $\alpha^2/N$ is small~($=5$), one can see a smooth transition in terms of $\Lambda$, but lying above the COE $\rightarrow$ CUE transition curve. The limit $\Lambda = 0$ corresponds to the TRI-preserved case, and $\Lambda=1$ to the completion of the TRI $\rightarrow$ TRNI transition. For large $\alpha^2/N \gtrsim 25$, the transition curves of $\Sigma^2(1,\Lambda)$ follow the COE $\rightarrow$ CUE transition. In Fig.~\ref{fig3}(b), we examine the transition curves for $\Sigma^2(2,\Lambda)$. We observe again that the transition in terms of $\Lambda$ is smooth. However, due to the larger interval size, the curves for small $\alpha^2/N$ are now more separated from the COE $\rightarrow$ CUE curve. For large $\alpha^2/N$, the transition curves again coincide with the COE $\rightarrow$ CUE curve. We have examined the transition curves of $\Sigma^2(r,\Lambda)$ in even larger intervals of size $r=5,10$ (not shown here). For a fixed $r$, the TRI $\rightarrow$ TRNI transition reduces to the COE $\rightarrow$ CUE transition in the limit $\alpha^2/N \gg r$. However, at a fixed value of $\alpha^2/N <N$, the transition starts deviating significantly from the COE $\rightarrow$ CUE transition for $r \sim \alpha^2/N$.

The contribution from the variance of significant perturbation matrix elements in $\Lambda$ is dependent on the parameter $\alpha^2/N$ and is large for small $\alpha^2/N$ (see Table~\ref{tab1}). Therefore, the required perturbation strength $\lambda$ for a fixed value of $\Lambda$ is smaller for smaller $\alpha^2/N$ --- see Eq.~\eqref{eq3}.

\begin{table}[t]
\caption{The perturbation strength $\lambda$ where $\Sigma^2(r,0)$ decays to $[\Sigma^2(r,0)+\Sigma^2(r,1)]/2$ for different values of $\alpha^2/N$ with $N$ = 2001.}
\label{tab2}
\begin{center}
\begin{tabular}{| c | c | c |} 
\hline
\multirow{2}{*}{\hspace{3mm}$\alpha^2/N$\hspace{3mm}} & \multicolumn{2}{c|}{$\lambda \times10^{5}$} \\
\cline{2-3}
& \hspace{3mm}$r$ = 1\hspace{3mm} & \hspace{3mm}$r$ = 2\hspace{3mm} \\
\hline
5 & 0.46 & 0.69 \\
\hline
10 & 0.83 & 1.04 \\
\hline
25 & 1.32 & 1.65 \\
\hline
50 & 1.85 & 2.32 \\
\hline
$\mathcal{O}(N)$ & 3.84 & 5.16 \\ [0.5ex]
\hline
\end{tabular}
\end{center}
\end{table}

Let us obtain a more quantitative understanding with the present scenario. We obtain the values of $\lambda$ where the number variance has decayed from $\Sigma^2(r,0)$ to $[\Sigma^2(r,0)+\Sigma^2(r,1)]/2$ for the data sets in Figs.~\ref{fig3}(a)-(b). In Table~\ref{tab2}, we list these values of $\lambda$ for different $\alpha^2/N$. For both $r=1$ and $2$, the value of $\lambda$ needed to drive transitions from $\Lambda=0$ to $\Lambda=1$ becomes less with decrease in $\alpha^2/N$. Therefore, we conclude that the TRI $\rightarrow$ TRNI transition in spectral fluctuations is enhanced when $\alpha^2/N$ is decreased below $N$.

\section{\label{s6} Eigenvector Fluctuations}

We next study the eigenvectors of the evolution operator. Since the transition in eigenvector fluctuations is also governed by $\Lambda$~\cite{PhysRevLett.54.2313,FRENCH1988198}, an enhancement in the TRI $\rightarrow$ TRNI transition should also show up in eigenvector fluctuations. We will work in a basis where the eigenvectors of $\mathcal{U}(\lambda=0)$ are real. Then, the speed of the TRI $\rightarrow$ TRNI transition describes the rate at which an eigenvector becomes complex with increase in $\lambda$ from $\lambda = 0$.

Let us consider the one-point eigenvector fluctuation measure, i.e., the probability distribution of the absolute square of an eigenvector component $y = |c_n|^2$, with $c_n$ denoting the $n\textsuperscript{th}$ component of an eigenvector. In RMT, the probability distribution function (normalized to unit mean) of $y$ has the form~\cite{RevModPhys.53.385,ROSENZWEIG1968437}
\be
\label{eq13}
P(y) = \frac{\left(\nu/2\right)^{1/2}}{\Gamma\left(\nu/2\right)}~y^{\nu/2-1} \rm{e}^{-\nu y/2},
\ee
where $\nu = 1$ for COE and 2 for CUE, respectively. The function $P(y)$ changes from a $\chi^2_1$-distribution with large variance ($\sigma^2 = 2$) to a $\chi^2_2$-distribution with smaller variance ($\sigma^2 = 1$) during the COE $\rightarrow$ CUE transition. The variance $\sigma^2$ in terms of $\Lambda$ continuously changes from $2$ to $1$ in this transition \cite{FRENCH1988198}. We stress that RMT results for eigenvector statistics in the crossover regime are still not fully available. There are a few preliminary works, but these do not precisely capture the intermediate regime (see~\cite{PhysRevE.49.R2513,PhysRevLett.85.2482} and references therein).

In this section, we numerically investigate the eigenvector fluctuations for different values of $\alpha^2/N$, where $N = 2001$. Note that the eigenvector statistics is basis-dependent. Therefore, we chose a generic basis, i.e., the position basis. We first calculate the eigenvectors of the evolution operator matrix $\mathcal{U}$ in Eq.~\eqref{eq7} using the standard LAPACK routines~\cite{laug}. Then, we find the absolute squares $|c_n|^2$ from all the eigenvectors of a single matrix and normalize them to mean unity. For eigenvector statistics, a single realization of the $\mathcal{U}$-matrix (yielding $2001\times2001$ eigenvector components) is sufficient to achieve good accuracy in data. To compare the probability distributions $P(y)$ for different $\alpha^2/N$, we plot $P(\log_{10}(y))$ vs. $\log_{10}(y)$. The RMT results for the COE $\rightarrow$ CUE transition are obtained numerically by using an interpolating ensemble of large Gaussian random matrices.

\begin{table}[t]
\caption{(i) Values of $\Lambda$ at $\lambda = 1.719\times10^{-5}$ for different values of $\alpha^2/N$ with $N=2001$. (ii) Corresponding values of $\sigma^2 (\lambda=0) - \sigma^2(\lambda=1.719\times10^{-5})$.}
\label{tab3}
\begin{center}
\begin{tabular}{| c | c | c |} 
\hline
\hspace{3mm}$\alpha^2/N$\hspace{3mm} & \hspace{3mm} $\Lambda$ \hspace{3mm} & \hspace{4mm} $\sigma^2(\lambda=0)-\sigma^2(\lambda=1.719\times10^{-5})$ \hspace{4mm} \\
\hline
5 & 0.25 & 0.92 \\
\hline
10 &  0.11 & 0.80 \\
\hline
25 & 0.05 & 0.52 \\
\hline
50 & 0.02 & 0.35 \\
\hline
$\mathcal{O}(N)$ & 0.005 & 0.12 \\ [0.5ex]
\hline
\end{tabular}
\end{center}
\end{table}

\begin{figure}[p]
\centering
\centering
\includegraphics[width=0.4\textwidth]{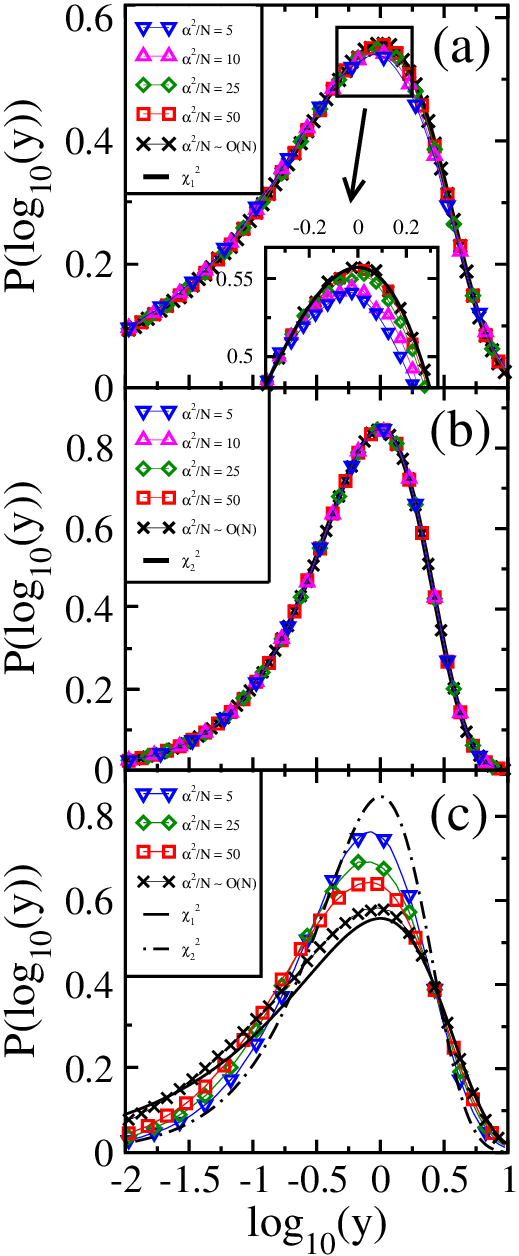}
\caption{Probability distribution $P(\log_{10}(y))$ vs. $\log_{10}(y)$ for various values of $\alpha^2/N$. The values of $\lambda$ are as follows: (a) $\lambda = 0$. The solid curve denotes the $\chi^2_1$-distribution. The inset magnifies the peak region. (b) $\lambda = 0.9$. The solid curve denotes the $\chi^2_2$-distribution. (c) $\lambda = 1.719 \times 10^{-5}$. The solid and dot-dashed curves denote the $\chi^2_1$ and $\chi^2_2$-distributions, respectively.}
\label{fig4}
\end{figure}

In Fig.~\ref{fig4}(a), we plot $P(\log_{10}(y))$ against $\log_{10}(y)$ for different $\alpha^2/N$ at $\lambda = 0$. The data sets for small $\alpha^2/N$~($= 5,10$) show a small departure from the $\chi^2_1$-distribution around the peak -- this region is amplified in the inset. On the other hand, for $\alpha^2/N \ge 25$, the data shows an excellent agreement with the $\chi^2_1$-distribution (COE). In Fig.~\ref{fig4}(b), we show data for different $\alpha^2/N$ at $\lambda = 0.9$. In this case, all data sets are in good agreement with the $\chi^2_2$-distribution (CUE). This shows that the eigenvectors become complex random states (irrespective of $\alpha^2/N$) when TRI is broken by a strong magnetic field. We stress that the one-point eigenvector fluctuations capture only the symmetries present in the evolution operator.

In Fig.~\ref{fig4}(c), we show the corresponding data for a small perturbation strength $\lambda = 1.719\times10^{-5}$. In this case, we expect an intermediate behavior between $\chi^2_1$ and $\chi^2_2$, depending on the value of $\Lambda$. Fig.~\ref{fig4}(c) shows that the data sets for smaller values of $\alpha^2/N$ approach closer to the limiting $\chi^2_2$-distribution. Therefore, the eigenvectors at large $\alpha^2/N$ require larger perturbations to reach the $\chi^2_2$ or CUE-limit. For the data in Fig.~\ref{fig4}(c), we present in Table~\ref{tab3} the following: (i) the values of the transition parameter $\Lambda$ from Eq.~\eqref{eq3}, and (ii) the difference between the variances of the distributions at $\lambda=0$ and $\lambda = 1.719\times10^{-5}$. For $\alpha^2/N = 5$, $\Lambda \simeq 0.25$ with a large decrease in $\sigma^2 \simeq 0.92$. For $\alpha^2/N \sim \mathcal{O}(N)$, $\Lambda \simeq 0.005$ and the change in $\sigma^2$ is relatively minor $\simeq 0.12$. This confirms that the eigenvector fluctuations for small $\alpha^2/N$ are more sensitive to $\lambda$. The reason behind this behavior is again the increased variance of perturbation matrix elements, which causes the strong admixing of imaginary amplitude in unperturbed (real) eigenvectors.

\begin{figure}[p]
\centering
\includegraphics[width=0.8\textwidth]{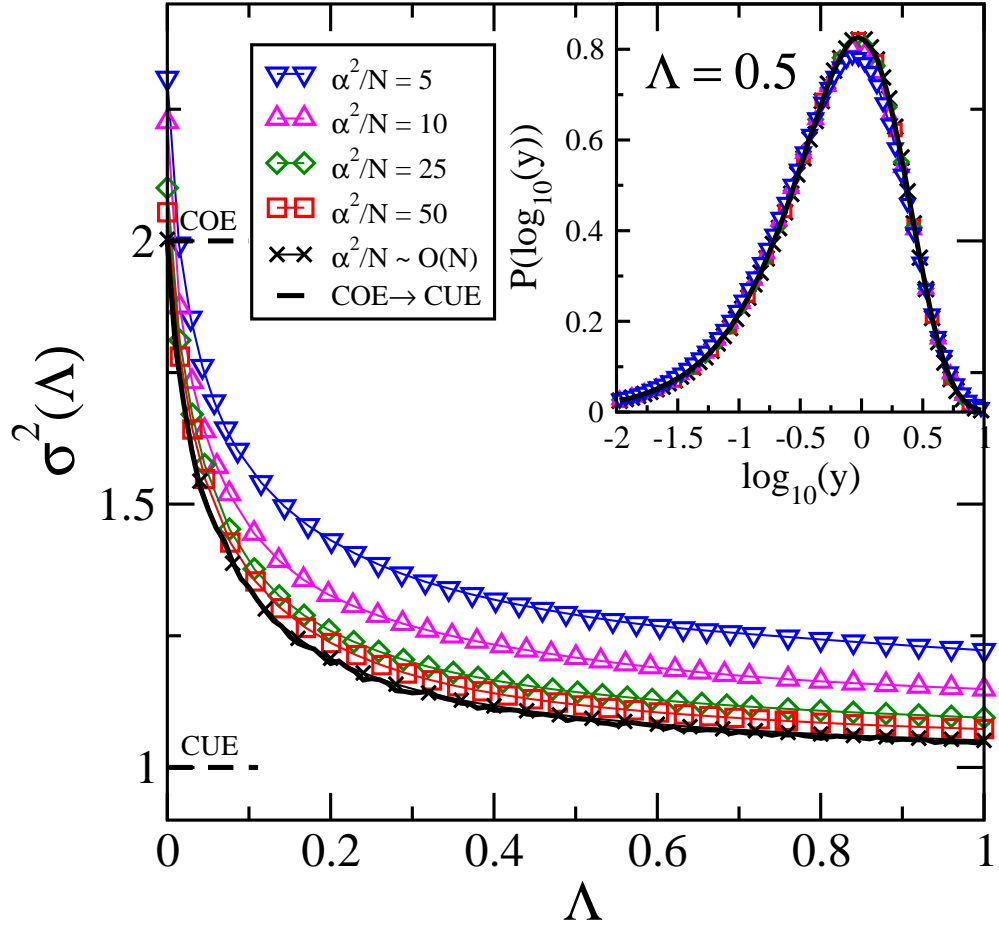}
\caption{Transition curves of the variance, $\sigma^2(\Lambda)$ vs. $\Lambda$, for various values of $\alpha^2/N$. The solid curve denotes the COE $\rightarrow$ CUE transition. The dashed lines indicate the COE and CUE limits. The inset plots $P(\log_{10}(y))$ vs. $\log_{10}(y)$ for the same values of $\alpha^2/N$, and a fixed value of $\Lambda = 0.5$. The solid curve in the inset is the distribution at $\Lambda = 0.5$ for the COE $\rightarrow$ CUE transition.}
\label{fig5}
\end{figure}

In Fig.~\ref{fig5}, we examine the transition in terms of $\Lambda$. We plot $\sigma^2$ vs. $\Lambda$ for different values of $\alpha^2/N$. The solid curve denotes the COE $\rightarrow$ CUE transition. The transition curves for $\alpha^2/N < N$ show a departure from the COE $\rightarrow$ CUE curve. (For $\alpha^2/N \sim \mathcal{O}(N)$, the transition curve is in good agreement with the COE $\rightarrow$ CUE curve.) Further, due to the increment in $v^2$, the speed of the transition is enhanced for small $\alpha^2/N$. In the inset of the same figure, we plot the distribution $P(\log_{10}(y))$ vs. $\log_{10}(y)$ for $\Lambda = 0.5$. For $\alpha^2/N \sim \mathcal{O}(N)$, the distribution is in very good agreement with the COE $\rightarrow$ CUE transition. As $\alpha^2/N$ is decreased below $N$, the distribution shows a deviation from the COE $\rightarrow$ CUE curve and becomes broader around the mean. This fact can be appreciated in the main frame, where $\sigma^2(\Lambda)$ is seen to be larger for smaller values of $\alpha^2/N$.

\section{\label{s7} Summary and Discussion}

Let us conclude this paper with a summary and discussion of our results. We have investigated the transition from \textit{time-reversal invariance} (TRI) to \textit{time-reversal non-invariance} (TRNI) in the spectral and eigenvector fluctuations of the \textit{quantum kicked rotor} (QKR) using Izrailev's finite-dimensional model. The local fluctuation properties of the QKR depend on the parameter $\alpha^2/N$, where $\alpha$ is the chaos parameter and $N$ is the dimensionality of the evolution operator matrix. The TRI in the QKR is broken immediately by the application of a small magnetic field. However, there is a continuous crossover from TRI $\rightarrow$ TRNI in the spectral and eigenvector fluctuations. This is governed by a local parameter $\Lambda$, which depends on the magnetic field and the near-diagonal matrix elements of the TRI-breaking perturbation. For $\alpha^2/N \gtrsim N$, the TRI $\rightarrow$ TRNI transition in the QKR is described well by the COE $\rightarrow$ CUE transition of \textit{random matrix theory} (RMT). This scenario applies if the TRI-breaking perturbation, when represented in the diagonal basis of the TRI-preserving part, becomes a member of Gaussian RMT ensembles.

However, for $\alpha^2/N < N$, the TRI-breaking perturbation no longer satisfies this. Rather, it has the form of a \textit{power-law random band matrix} (PRBM), whose bandwidth grows linearly with $\alpha^2/N$. The power-law fall of the variance of perturbation matrix elements outside the band has a decay exponent $m \simeq 1.35$, which is the same for all values of $\alpha^2/N < N$. (This value of $m$ is reminiscent of so-called sub-critical PRBM models~($1<m<2$)~\cite{PhysRevE.54.3221,PhysRevB.82.125106}.)

In addition to the PRBM structure, $Var(1)$ increases linearly with $N/\alpha^2$. The increased variance for small $\alpha^2/N$ enhances the sensitivity of $\Lambda$ to the perturbation strength, resulting in a faster TRI $\rightarrow$ TRNI transition as a function of the magnetic field.

For spectral fluctuations, we investigated the transition curves of rigidity measure [i.e., number variance $\Sigma^2(r,\Lambda)$, where $r$ is the size of the spectral interval] by varying $\Lambda$ at fixed $r$. For small $\alpha^2/N$, the transition curve deviates from the COE $\rightarrow$ CUE transition. Further, the speed of this transition is enhanced for smaller values of $\alpha^2/N$. As we increase $\alpha^2/N$, the COE $\rightarrow$ CUE transition is approached. The eigenvector fluctuations are studied via the probability distributions $P(y)$ of the absolute square of the eigenvector component $y$. For a small increment in $\lambda$, the variance $\sigma^2$ of $P(y)$ changes substantially for small $\alpha^2/N$. We also analyzed the transition of variance $\sigma^2$ with respect to the parameter $\Lambda$. For $\alpha^2/N < N$, the transition curve of $\sigma^2$ deviates from the COE $\rightarrow$ CUE curve. For $\alpha^2/N \sim \mathcal{O}(N)$, the TRI $\rightarrow$ TRNI transition in eigenvectors is consistent with the COE $\rightarrow$ CUE transition. Again, the speed of the transition in the eigenvector fluctuations is higher for smaller values of $\alpha^2/N$ below $N$.

A similar enhancement in the TRI $\rightarrow$ TRNI transition was predicted by French et al.~\cite{PhysRevLett.58.2400,FRENCH1988235} in the context of a nuclear Hamiltonian. These authors modeled the TRI-preserving and TRI-breaking parts of the Hamiltonian as $m$-particle embedded-GOE ensembles with $k$-body interactions. In the limit $m \gg k$, the near-diagonal variance of the perturbation matrix was seen to increase. This resulted in an enhancement of the transition. However, a semi-classical treatment of this enhanced transition remains an open problem.

\begin{acknowledgments}
We would like to thank Avanish Kumar for discussions.
\end{acknowledgments}

\newpage
\appendix

\section{Time-evolution Operator $\mathcal{U}$ in Infinite-dimensional Position Basis}
\label{infinite}

Consider the form of the operator $\mathcal{U}$ in Eq.~\eqref{eq6}. The matrix elements of $\mathcal{U}$ in position basis $\{\vert q \rangle : \theta \vert q \rangle = \theta_q \vert q \rangle\}$ are given by
\be
\label{eqAA1}
\mathcal{U}_{qq^{\prime}} = \langle q \vert B^{1/2}GB^{1/2} \vert q^{\prime} \rangle.
\ee
The position basis for the QKR is periodic $(\vert q + 2\pi \rangle = \vert q \rangle)$. Then, the orthonormality and completeness relations in the position basis $\{\vert q \rangle\}$ can be defined as~\cite{albert2017,PhysRevA.81.052327}
\be
\label{eqAA2}
\langle q \vert q^{\prime} \rangle = 2\pi ~ \delta^{(2\pi)} (q-q^{\prime}), \quad \mbox{and} \quad
\int_{(2\pi)} \frac{d q}{2\pi} ~ \vert q \rangle \langle q \vert = \mathbb{I},
\ee
where the superscript $(2\pi)$ on the $\delta$-function indicates $2\pi$-periodicity. The integration for variable $q$ is over a $2 \pi$ interval.

Due to the periodicity of the position basis, the eigenvalues of the momentum operator are discrete, i.e., $p \vert l \rangle = l \hbar \vert l \rangle$; $l = 0, \pm 1, \pm 2, \ldots$. The orthonormality and completeness relations in the momentum basis are expressed as
\be
\label{eqAA3}
\langle l \vert l^{\prime} \rangle = \delta_{l,l^{\prime}}, \quad \mbox{and} \quad
\sum_{l=-\infty}^{l=\infty} \vert l \rangle \langle l \vert = \mathbb{I}.
\ee
Further, the inner products between the basis vectors of the discrete momentum basis and the continuous position basis are defined as
\be
\label{eqAA4}
\langle q \vert l \rangle = {\rm e}^{i l \theta_q}, \quad \langle l \vert q \rangle = {\rm e}^{-i l \theta_q}.
\ee

By using the completeness relations from Eqs.~\eqref{eqAA2}--\eqref{eqAA3}, the matrix element $\mathcal{U}_{qq^{\prime}}$ in Eq.~\eqref{eqAA1} can be written as
\be
\label{eqAA5}
\mathcal{U}_{qq^{\prime}} = \frac{1}{(2 \pi)^2} \sum_{l,l^{\prime} = -\infty}^{\infty} \int_{(2\pi)} \int_{(2\pi)} dr ~ dr^{\prime} ~\langle q \vert B^{1/2} \vert r \rangle~\langle r \vert  l \rangle~\langle l \vert G \vert l^{\prime} \rangle~\langle l^{\prime} \vert r^{\prime}  \rangle~\langle r^{\prime} \vert B^{1/2}\vert q^{\prime} \rangle.
\ee
Here, the continuous variables $q$, $q^{\prime}$, $r$ and $r^{\prime}$ label the position basis, and the integers $l$ and $l^{\prime}$ label the momentum basis.

By using Eq.~\eqref{eqG} for $G$ in Eq.~\eqref{eqAA5}, and simplifying the expression using  Eqs.~\eqref{eqAA2}--\eqref{eqAA4}, we obtain
\be
\mathcal{U}_{qq^{\prime}} = B^{1/2}(\theta_q) B^{1/2}(\theta_q^{\prime}) \sum_{l=-\infty}^{l=\infty} {\rm e}^{-i T (l \hbar - \lambda)^2/(2\hbar)} ~ {\rm e}^{i l (\theta_q-\theta_q^{\prime})}.
\ee
Further, recalling the definition of $B$ from Eq.~\eqref{eqB}, the matrix elements $\mathcal{U}_{qq^{\prime}}$ take the form given in Eq.~\eqref{eqInf} of the main text.

\newpage
\bibliography{qkrenh.bib}

\end{document}